\documentclass{article}

\usepackage{PRIMEarxiv}
\usepackage{amsmath} 

\usepackage[utf8]{inputenc} 
\usepackage[T1]{fontenc}    
\usepackage{hyperref}       
\usepackage{url}            
\usepackage{booktabs}       
\usepackage{amsfonts}       
\usepackage{nicefrac}       
\usepackage{microtype}      
\usepackage{lipsum}
\usepackage{fancyhdr}       
\usepackage{graphicx}       
\graphicspath{{media/}}     
\usepackage{subcaption}
\usepackage{multirow} 

\title{Mixture of Experts with Mixture of Precisions for Tuning Quality of Service
}

\author{
  HamidReza Imani, Abdolah Amirany, and Tarek El-Ghazawi \\
  Department of Electrical and Computer Engineering\\
  The George Washington University \\
  Washington, DC, USA\\
  \texttt{\{hamidreza, a.amirany, tarek\}@gwu.edu} \\
}

\begin{document}
\maketitle

\begin{abstract}
The increasing demand for deploying large Mixture-of-Experts (MoE) models in resource-constrained environments necessitates efficient approaches to address their high memory and computational requirements challenges. Moreover, given that tasks come in different user-defined constraints and the available resources change over time in multi-tenant environments, it is necessary to design an approach which provides a flexible configuration space. This paper presents an adaptive serving approach for the efficient deployment of MoE models, capitalizing on partial quantization of the experts. By dynamically determining the number of quantized experts and their distribution across CPU and GPU, our approach explores the Pareto frontier and offers a fine-grained range of configurations for tuning throughput and model quality. Our evaluation on an NVIDIA A100 GPU using a Mixtral 8x7B MoE model for three language modelling benchmarks demonstrates that the throughput of token generation can be adjusted from 0.63 to 13.00 token per second. This enhancement comes with a marginal perplexity increase of 3.81 to 4.00, 13.59 to 14.17, and 7.24 to 7.40 for WikiText2, PTB, and C4 datasets respectively under maximum quantization. These results highlight the practical applicability of our approach in dynamic and accuracy-sensitive applications where both memory usage and output quality are important.
\end{abstract}


\section{Introduction}

Utilizing Mixture-of-Experts (MoE) architectures \cite{jacobs1991adaptive, jordan1994hierarchical} in Large Language Models (LLMs) has significantly enhanced the performance of Natural Language Processing (NLP) tasks \cite{shazeer2017outrageously, shazeer2018mesh}. MoE-based transformer models employ multiple parallel feed-forward (FF) layers, known as experts, within each transformer block instead of a single FF layer. In this architecture, a gating network is used to evaluate the input token generated by the attention layer and to assign weights to each expert to perform a weighted averaging. As a result, by scaling and adding more parameters, MoE models are able to achieve superior performance comparing to their dense counterparts.

With parameters reaching into the trillions, MoE models can expand to terabyte scale, making them challenging to deploy for inference. Most of the available GPUs do not have sufficient resources to store the full model on their main memory. Therefore previously proposed deployment approaches load the MoE model on GPU partially and consider a swap space for the experts to be transferred between memories of CPU and GPU during inference which causes recurring data movement and copy operations. Since these data movements take more time comparing to the actual computation on the GPU, they become bottleneck in the inference pipeline.


Meanwhile, there is an increasing demand for deployment of customized LLMs in companies and research labs without access to abundant resources or high-end GPUs. Additionally, in real-life scenarios, the computing systems available to these entities are often shared among multiple users, with available resources fluctuating and user constraints changing over time. Therefore, it is important to design an adaptive serving systems that can accommodate deployment of large MoE models in dynamic resource-constrained settings.

Prior work has covered different aspects to alleviate the inference process of MoE models. Model compression techniques such as \cite{eliseev2023fast, frantar2023qmoe, kim2022says, kim2023mixture} perform post-training quantization to directly reduce the size which can mitigate CPU-GPU communication overhead. However, these approaches reduce the quality of the output generated text considerably. Meanwhile, efficient serving approaches such as \cite{eliseev2023fast, xue2024moe} treat the memory available on the GPU as a cache and each expert of the MoE model as a cache block therefore trying to minimize data movements by increasing the hit rate. Although these approaches improve the performance of MoE inference in throughput and output quality, they do not provide the required flexibility to handle variable constraints. 

In this paper, we focus on dynamic single-GPU settings and propose partial quantization of experts in MoE models. Since the main bottleneck of MoE deployment is data transfers between GPU and CPU and since the experts constitute most of the model size, we only focus on experts to reduce the model's overall memory footprint. Moreover, to provide flexibility regarding to user-specific needs (task preference) and the available memory, we adaptively determines the number of 16-bit and 4-bit experts and their partitioning among CPU and GPU. This allows us to have fine-grained control over the quality of the generated output, and throughput of token generation making it suitable for dynamic and accuracy sensitive applications. We evaluate our adaptive serving approach on an NVIDIA A100 GPU using a Mixtral 8x7B MoE model \cite{jiang2024mixtral}. The results show that the proposed approach is able to improve the throughput significantly while imposing negligible perplexity increase.

\section{Related Work}

\subsection{Mixture-of-Experts}

The MoE architecture was first proposed in \cite{jacobs1991adaptive} to allow each FF network to specialize in a subset of the input space when dealing with clustered data. Later, in \cite{shazeer2017outrageously}, the MoE architecture was employed in a machine translation task to insert experts between stacked LSTM layers. 

To apply the MoE concept to Transformer models, in \cite{fedus2022switch, du2022glam, kim2021scalable, zuo2021taming, lin2021m6, rajbhandari2022deepspeed}, the dense FF layers are replaced by sparse switch FF layers in the Transformer block. To make these models computationally efficient, a top-k sparse gating mechanism is employed which selects only a subset of experts in each layer. This approach enables MoE models to scale linearly in size while imposing a minimal increase in computation, depending on the number of experts included in weighted averaging. This also makes training faster and more efficient compared to a dense model with the same number of parameters, since gradients are applied sparsely to the activated experts in the forward pass \cite{liu2023sparse}.

\subsection{Efficient Serving and Orchestration of MoE Models}
To serve the MoE models efficiently, it is important to keep the model parameters as close to the GPU as possible. However, the model's large sizes make it impossible to store all the parameters on the limited accelerator memory. Therefore, as proposed in \cite{sheng2023flexgen}, maximum number of model parameters are stored on GPU and the remainder of them are transferred between GPU and CPU during inference. 

To reduce the communication overhead caused by weight transfers, authors in \cite{kamahori2024fiddler}, propose a dual-track execution flow. In cases where the selected expert is not available on GPU, their approach transfers the activations from GPU to CPU instead of bringing the expert to the GPU. This method allows computations for that specific expert to be performed on a multi-core CPU, resulting in higher overall throughput. However, this approach necessitates a powerful many-core CPU to offset the expert transfer overhead.

Eliseev et. al. \cite{eliseev2023fast} utilize a table for tracking the experts and using the least recently used (LRU) policy they replace the experts on GPU. Due to residual connections from each transformer block to the next, the authors also heuristically feed the same activation from previous layers to the gating network of subsequent layers to predict potential experts for speculative prefetching.

In \cite{xue2024moe}, authors preprocess a specific dataset and design a data structure (table) to capture access patterns of experts for each prompt. All distinct access patterns are stored in a library, and during inference, a number of the most similar tables are selected based on accesses to experts in earlier layers, to determine which experts to prefetch.

\subsection{Model Compression and Quantization}

Model compression and quantization are highly effective approaches for efficient deployment of LLMs. These approaches can be categorized into two major groups: quantization aware training (QAT) and post training quantization (PTQ). QAT approaches \cite{liu2023llm, dettmers2024qlora} refine and retrain the quantized model iteratively to improve the accuracy and therefore are task-specific. Meanwhile, PTQ approaches \cite{lin2023awq, xiao2023smoothquant, huang2024billm, eliseev2023fast, dettmers2023case} can be applied to any pretrained model in a single shot which impose less overhead for compression.

In another direction, more specific to MoE models, He et. al. \cite{he2024demystifying} explore model compression through pruning and dropping the model parameters. Authors explore the decrease in the output quality of a Mixtral MoE model caused by expert compression (reducing size of an expert using quantization or pruning), MoE Layer Drop (removing all the experts from a transformer block), and Block Drop (removing the whole transformer block). Unlike the PTQ and QAT, the approach proposed by He et al. \cite{he2024demystifying} does not transform the full-precision model into a fixed-size quantized model; instead, it provides the flexibility to choose the model size.

\section{Adaptive Inference Partitioner and Planner}
\begin{figure*}
  \centering
  \includegraphics[width=0.9\linewidth]{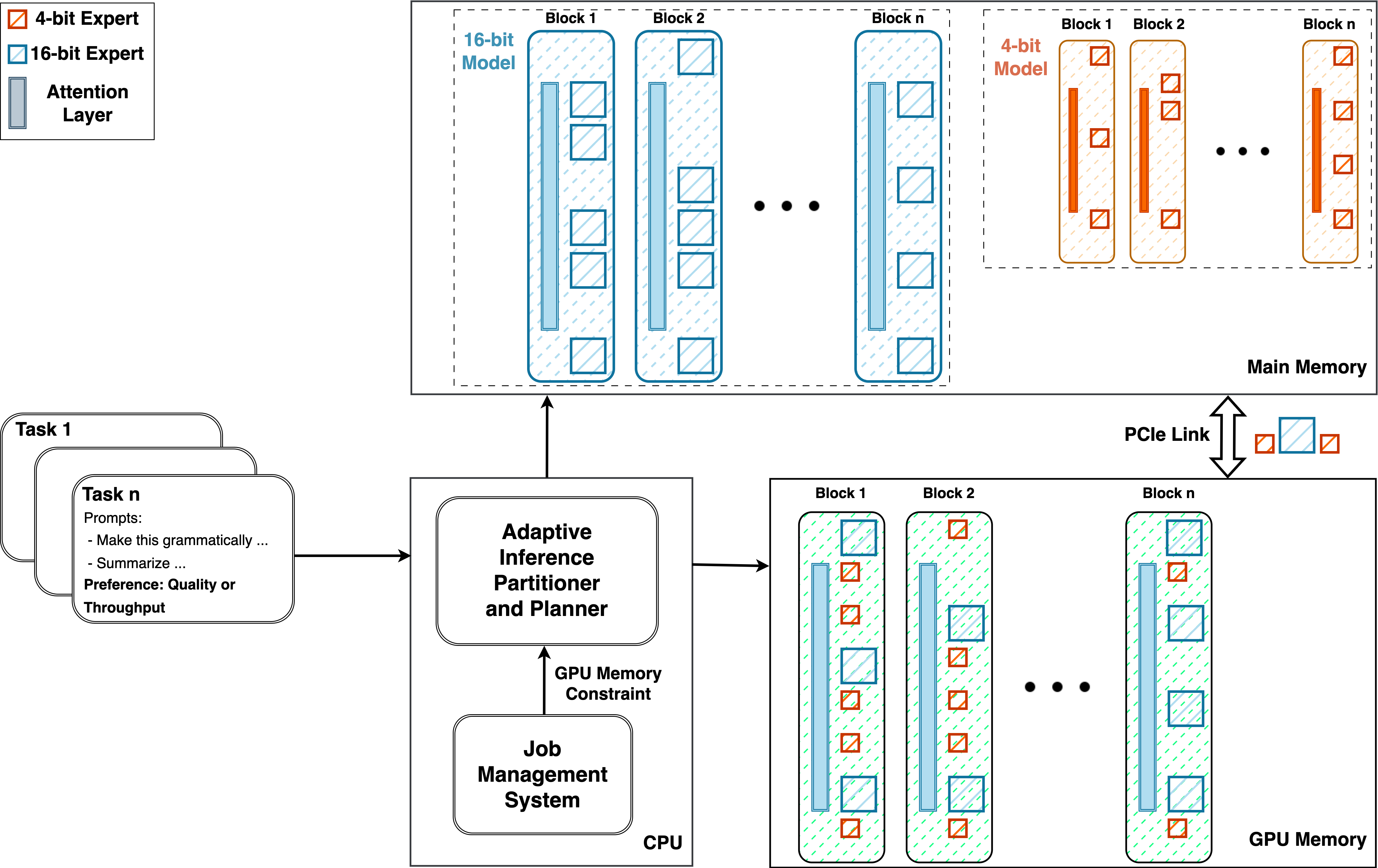}
  \caption{An adaptive inference partitioner and planner for deployment of MoE models}
  \label{adaptive_planner}
\end{figure*}

The motivation behind this partitioning and planning system is to adapt to the requirements of a multi-tenant environment. In these environments, resources are shared among users, and job management system determines the amount of resources (GPU utilization, PCIe link bandwidth, and GPU memory) available at each point. Since these constraints change over time, an adaptive system is required to adjust the configuration of the inference system accordingly. Here, we focus the scope of the design on the case where the amount of available GPU memory imposes constraints on the system.

In this design, since all the non-expert layers together constitute a small portion of the model size (only 5\% for the Mixtral-8x7B model) and have a significant effect on the output quality, we load them onto the GPU in 16-bit format. Therefore, to meet the memory constraint, the partitioning system offers two options to control the amount of allocated memory with higher granularity: 1) partial expert quantization, which determines the number of quantized vs. 16-bit experts, and 2) deciding the number of experts that reside on the GPU and the number of experts that are transferred during inference.

As depicted in Figure \ref{adaptive_planner}, we assume that tasks arrive at the inference system with a batch of prompts and a preference for either quality or throughput. Based on the task's preference, one of the options mentioned above receives higher priority. If the preference is for throughput, the system must bring as many experts as possible into the GPU memory to reduce the amount of data transfers caused by an expert miss. Therefore, if the available memory (denoted as $Mem_{GPU}$) exceeds the total size of non-expert layers (denoted as $Size_{N.E.}$) in 16-bit and experts in 4-bit (denoted as $Num_E \times Size_{E,4}$), some of the experts will be selected to stay in 16-bit format (denoted by $Num_{E,16}$ and calculated in equation \ref{eq1}) to utilize the excess resources for better quality. On the other hand, when the available memory is less, all of the experts will be assigned to the 4-bit format, and the system will have to perform offloading.

\begin{equation} \label{eq1}
Num_{E,16} = 
\begin{cases}
    \lfloor(Mem_{GPU} - Size_{N.E.} - \\ Num_E  \times Size_{E,4})/(3 \times Size_{E,4}) \rfloor & \text{if } Mem_{GPU} > Size_{N.E.} + Num_E \times Size_{E,4} \\
    0 & \text{otherwise}
\end{cases}
\end{equation}

Meanwhile, for tasks with a quality constraint, a range for selecting the number of quantized experts is provided. The upper bound of this range achieves the best output quality by setting all the experts to perform in 16-bit, while the lower bound keeps all the experts quantized. Therefore, the number of 4-bit experts ($Num_{E,4}$), along with the available GPU memory, directs the partitioner to distribute the model across CPU and GPU memories. If the available GPU memory is less than the size of the non-expert layers and $Num_{E,4}$ 4-bit experts, only a subset of quantized experts will be stored on the GPU, and the rest of the experts will be transferred during inference. On the other hand, with a higher amount of available memory, more experts will be loaded onto the GPU, which does not change the output quality but can improve the throughput.

Based on the partitioning mentioned above, we consider a table that has two boolean attributes for each expert. The first attribute determines if the expert is quantized or not, and the second attribute determines the location of the expert (CPU or GPU). First, the quantization attribute is assigned to experts randomly. This approach ensures that the method is not dependent on a specific dataset or task since MoE models are trained to have uniform access frequency among all experts. Next, the location attribute is assigned. Since an expert not being available on the GPU stalls the computations in the pipeline, we give higher priority to 4-bit experts and assign them to the GPU first to increase the hit rate of the inference process.

For inference, the model is partially loaded onto the GPU, and the state of each expert is tracked using the described table. Additionally, a swap space is allocated to transfer experts from the CPU when an expert miss occurs. Since user constraints can change over time, the system is designed to adapt with minimal downtime. The planner recalculates the parameters based on the new constraints and partially reconfigures the system instead of reloading the model with the new configuration as a whole. These partial reconfigurations involve offloading or downloading experts and switching between quantized and 16-bit formats.





\section{Evaluation}

\subsection{Experimental Setup}
We develop and deploy our adaptive serving approach using Pytorch \cite{paszke2019pytorch} and Huggingface \cite{wolf2019huggingface} libraries. Also, we use the Bitsandbytes library \cite{dettmers2023case} for quantization of experts. We conduct our experiments on a server which has an AMD 16-Core MILAN CPU and a single 80GB NVIDIA A100 GPU with PCIe Gen4 interconnect .

\textbf{Model and Datasets.} We evaluate our approach using a Mixtral 8x7B MoE model \cite{jiang2024mixtral}, which consists of 32 layers and 8 experts per layer. The total size of the non-expert layers for this model adds up to 3.16 GB. Each expert occupies 336 MB of memory, and transferring each expert between the CPU and GPU takes 27.35 ms on the described testbed.

To assess the effect of partial expert quantization in our approach, we evaluate the performance of the MoE model using three language modeling benchmarks. We examine the model on three popular text datasets: WikiText2 \cite{merity2016pointer}, PTB \cite{marcus1994penn}, and C4 \cite{raffel2020exploring}. For each dataset, we measure the perplexity of the model's generated output using 128 samples of 2048 tokens, as perplexity is known as a robust and widely adopted indicator of model capabilities \cite{huang2024billm}.

\textbf{Baselines.} Since our adaptive serving approach provides a design space for choosing between model quality and throughput, our comparison against baseline approaches focuses on these two metrics. For perplexity, we compare our approach's performance against models with homogeneous weight formats: 16-bit, 8-bit, and 4-bit \cite{dettmers2023case}. Regarding the efficiency of token generation, we compare our approach's throughput (tokens per second) against Mixtral-Offloading \cite{eliseev2023fast} under varying amounts of available GPU memory. The throughput is measured using a mix of prompts from the mentioned datasets, with input and output lengths limited to 16 to balance the prefill and decode stages' latencies.

\subsection{Results}

\begin{figure*}[ht]
    \centering
    \begin{subfigure}[t]{0.345\textwidth} 
        \centering
        \includegraphics[width=\linewidth]{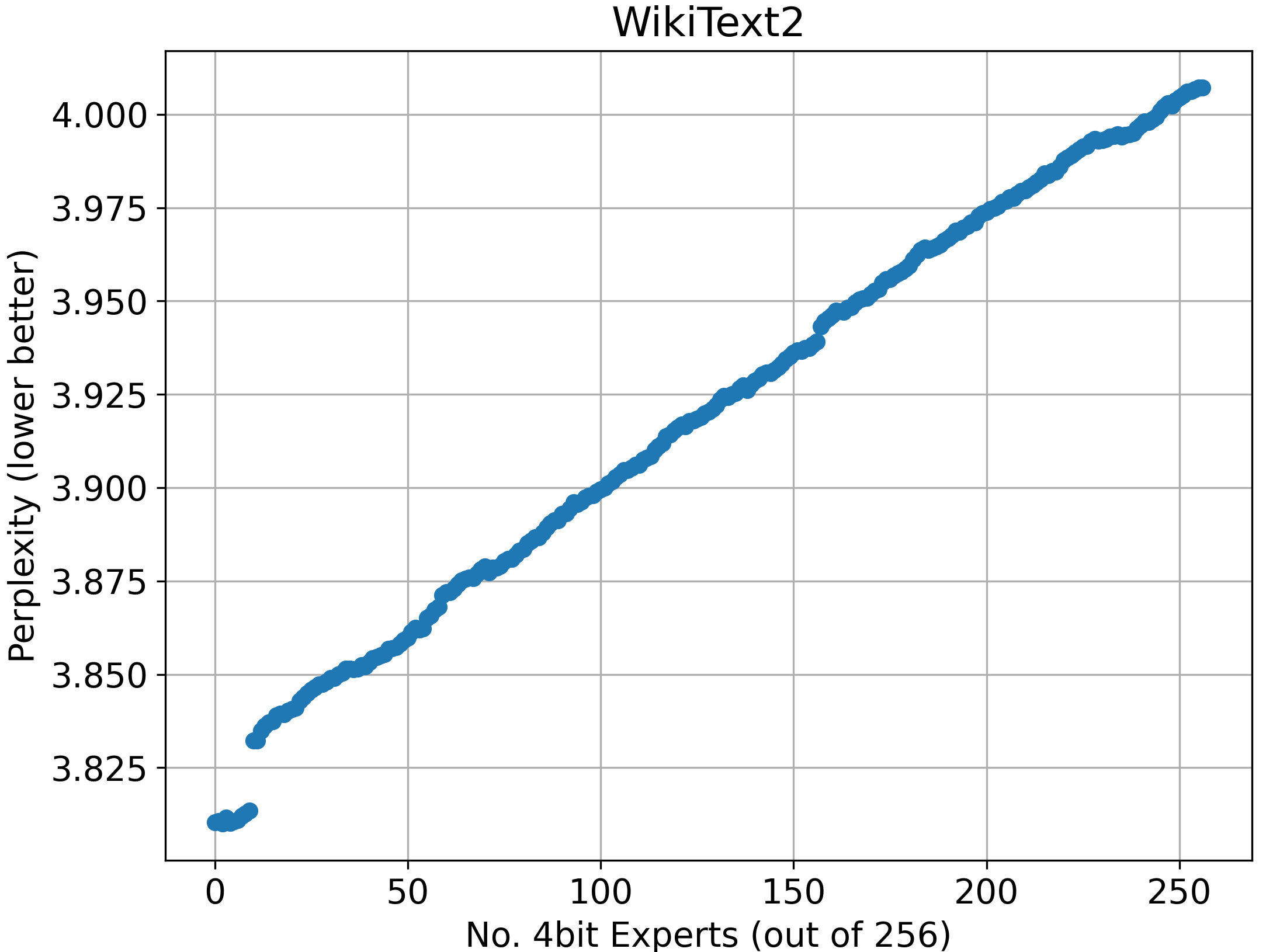}
    \end{subfigure}%
    \begin{subfigure}[t]{0.325\textwidth} 
        \centering
        \includegraphics[width=\linewidth]{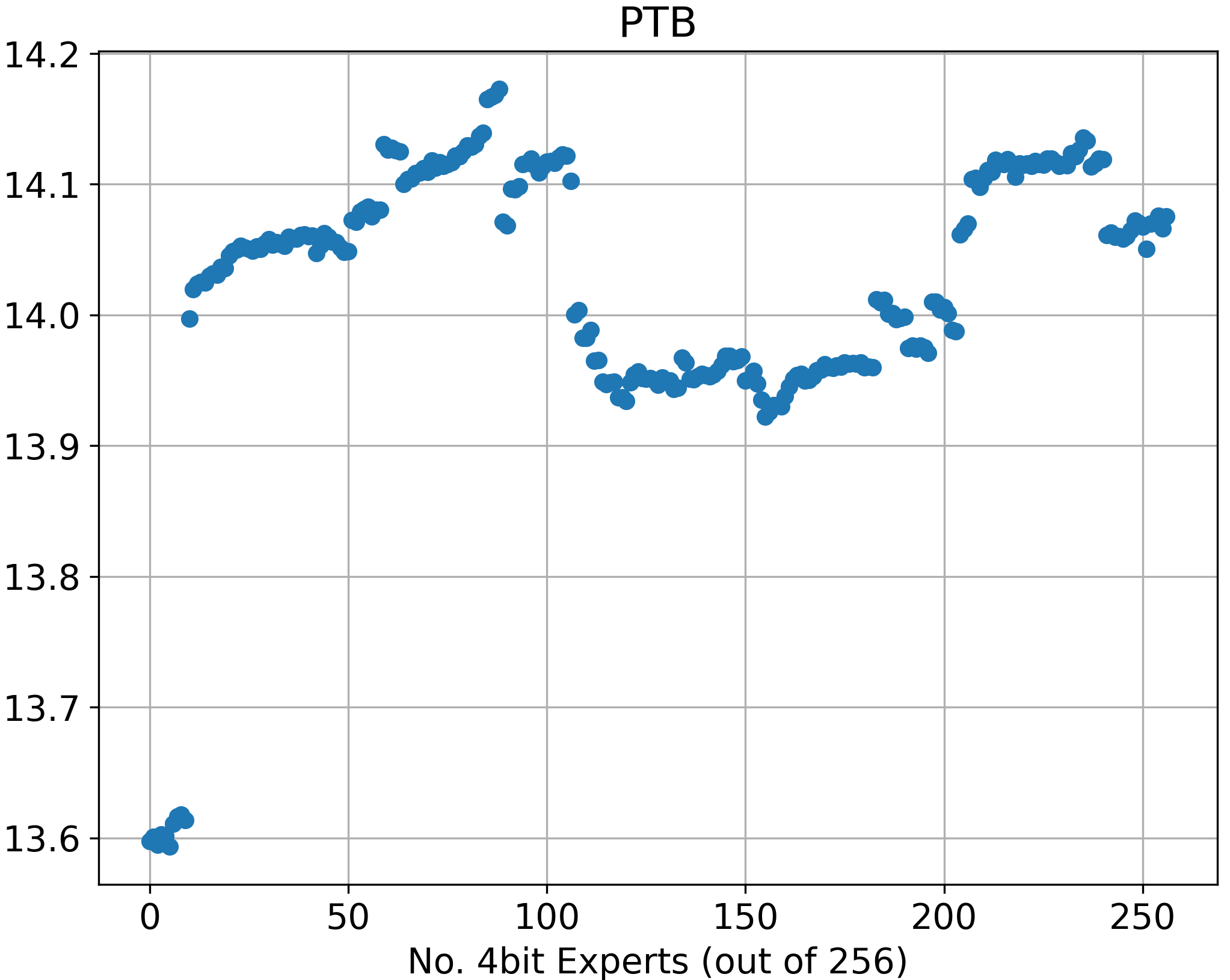}
    \end{subfigure}%
    \begin{subfigure}[t]{0.325\textwidth} 
        \centering
        \includegraphics[width=\linewidth]{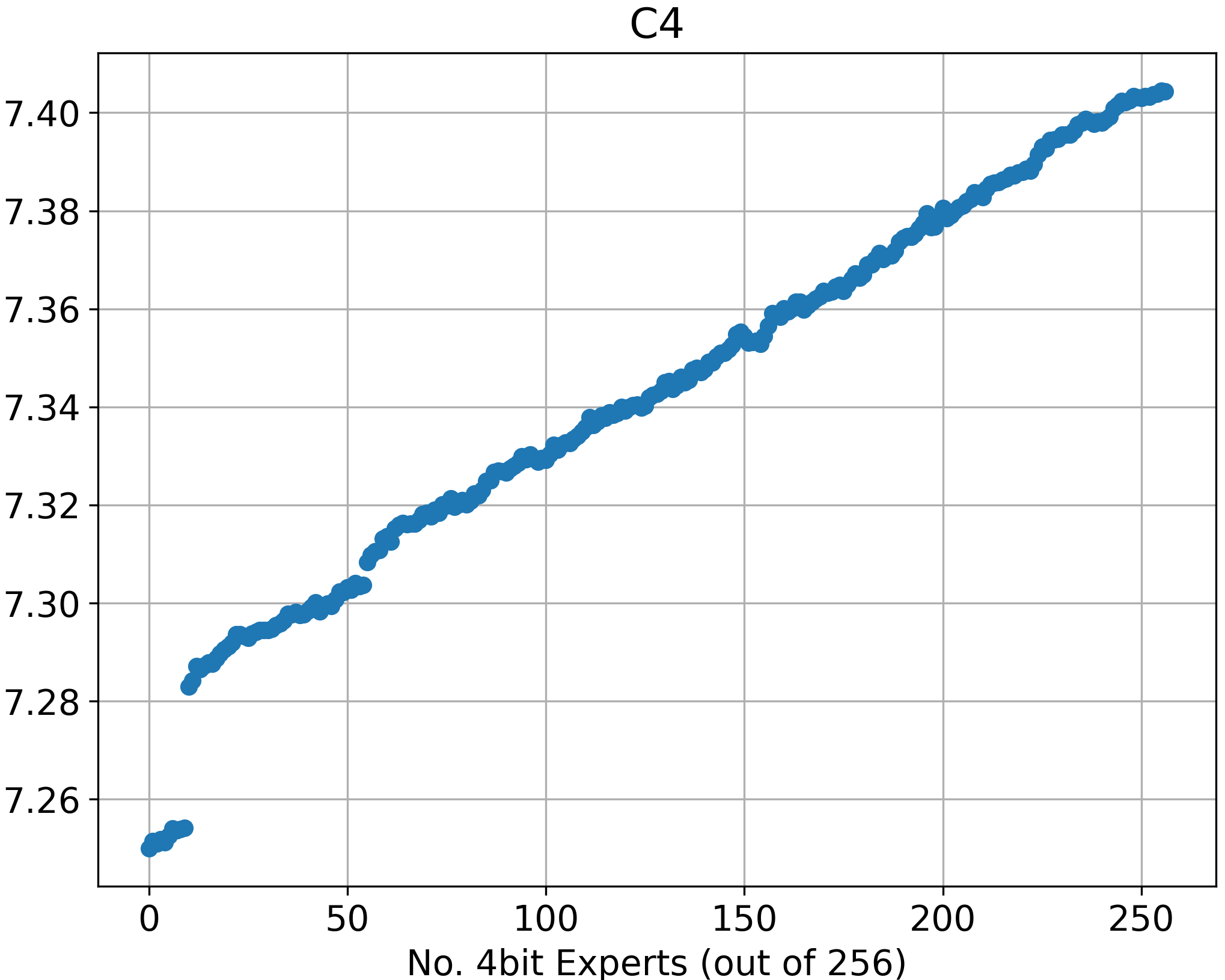}
    \end{subfigure}
    
    \caption{Perplexity of the expert-only partially quantized model across varying numbers of 4-bit experts (out of a total of 256 experts) }
    \label{perplexity}
\end{figure*}

Here, we first evaluate the effect of quantized experts on the performance of the MoE model. As shown in Figure \ref{perplexity}, increasing the number of 4-bit experts generally leads to a decrease in the quality of the generated output (higher perplexity), although this trend is not always strictly monotonic. In the case of the PTB dataset, there are points in the graph where increasing the number of quantized experts decreases the perplexity. This might be due to the perplexity metric's limitation in accurately evaluating minor modifications to the MoE model and also quantization smoothing the decision boundaries of the model \cite{ansari2019improving}. 

\begin{figure*}[ht]
    \centering
    \begin{subfigure}[t]{0.46\textwidth} 
        \centering
        \includegraphics[width=\linewidth]{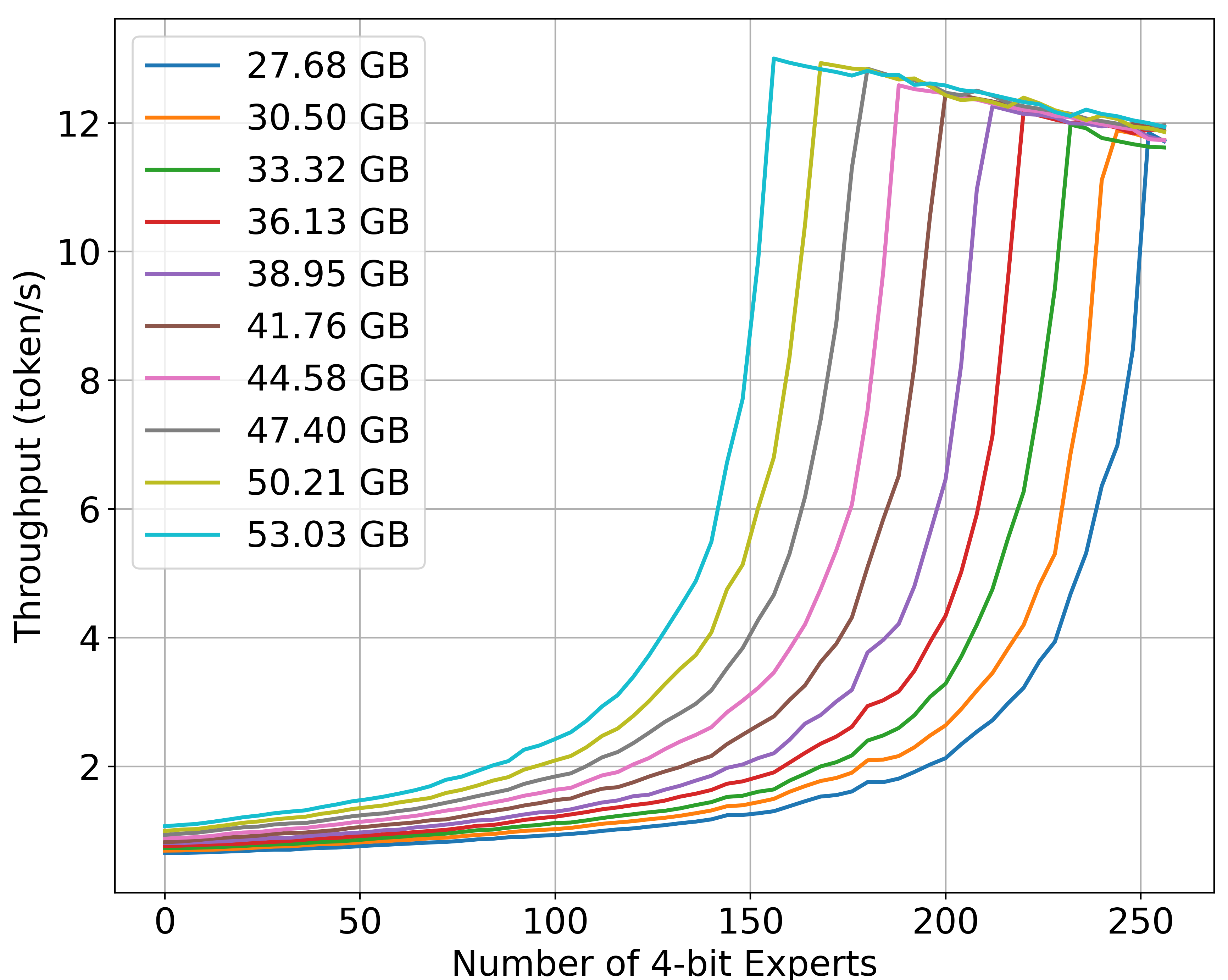}
    \end{subfigure}%
    \begin{subfigure}[t]{0.53\textwidth} 
        \centering
        \includegraphics[width=\linewidth]{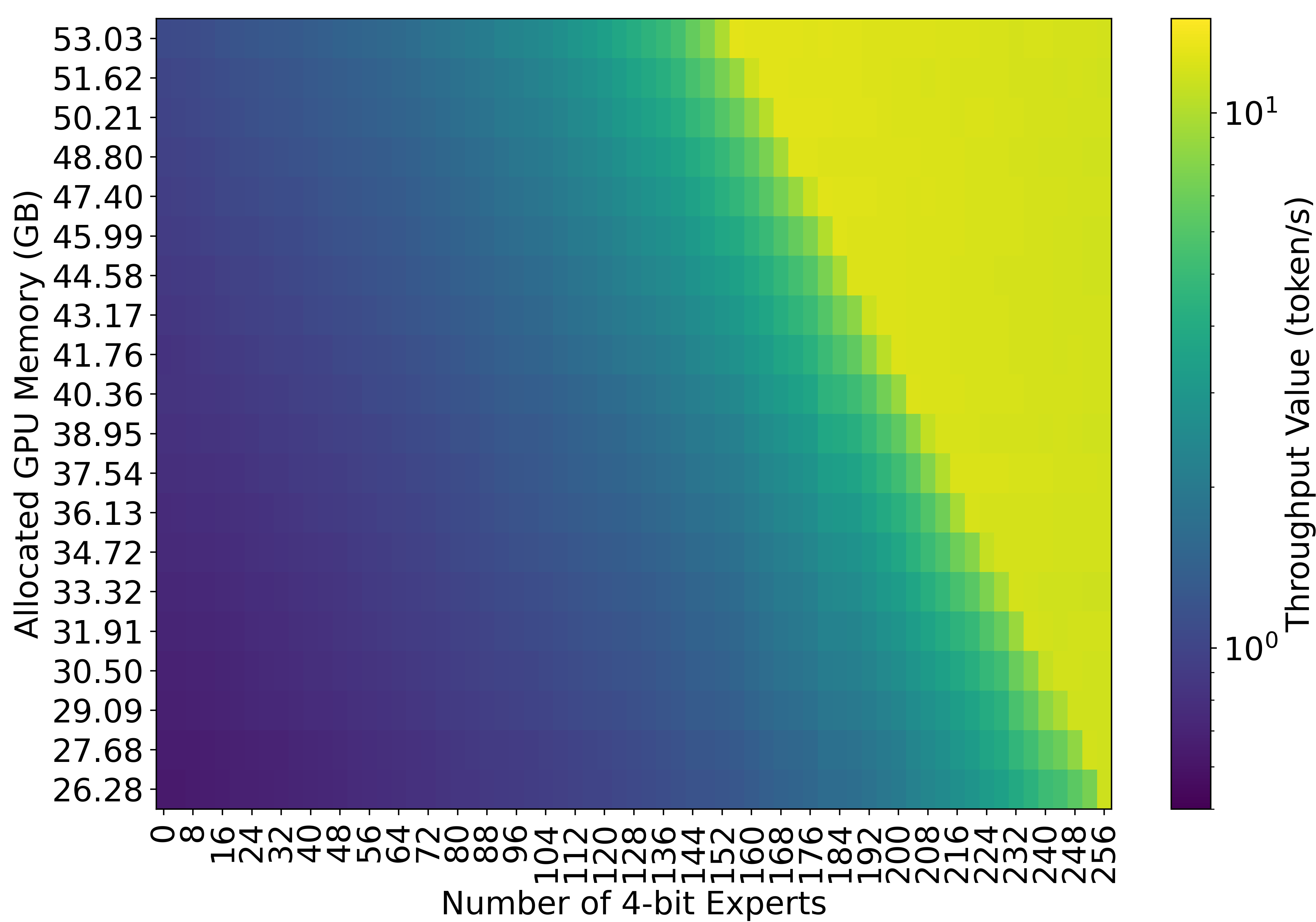}
    \end{subfigure}%

    \caption{Throughput of an expert-only partially quantized Mixtral 8x7B MoE model running on an NVIDIA A100 GPU under different amounts of available memory}
    \label{throughput}
\end{figure*}

In Figure \ref{throughput}, we evaluate the throughput of the partially quantized model under varying amounts of available GPU memory. The yellow triangle on the right side of the graph marks the region where all parameters of the partially quantized model are stored on the GPU, resulting in maximum throughput. Meanwhile, increasing the number of quantized experts in this region causes a slight drop in throughput, due to PyTorch's slower 4-bit matrix multiplication compared to the 16-bit counterpart. Conversely, in the region where offloading is required, throughput decreases due to communication overhead. Moreover, in this region, as both available memory and the number of quantized experts increase (which can be seen as mitigation of model memory footprint), throughput exhibits hyperbolic growth.

Furthermore, as summarized in Table \ref{comparison}, we compare the overall memory footprint and the performance of the language models across different quantization configurations. Comparing the partially quantized MoE model with the homogeneous 8-bit model demonstrates that the adaptive serving approach offers a design space that can both enhance the model's output quality and reduce the model size. Moreover, the maximum value of perplexity in partial quantization shows that the quantization effect of expert layers on perplexity is negligible compared to that of non-expert layers, although it significantly helps in reducing the model size.

\begin{table*}
\centering
\begin{tabular}{cccccc}
\hline
\multirow{2}{*}{\textbf{Non-Expert Quant.}} & \multirow{2}{*}{\textbf{Expert Quant.}} & \multirow{2}{*}{\textbf{Model Size (GB)}} & \multicolumn{3}{c}{\textbf{Perplexity}}         \\ \cline{4-6} 
                                           &                                        &                                           & \textbf{WikiText2} & \textbf{PTB} & \textbf{C4} \\ \hline
4-bit                                      & 4-bit                                  & 23.55                                     & 4.20               & 14.38        & 7.61        \\ \hline
8-bit                                      & 8-bit                                  & 47.10                                     & 3.82               & 13.25        & 7.26        \\ \hline
16-bit                                     & 16-bit                                 & 94.21                                     & 3.81               & 13.59         & 7.24        \\ \hline
16-bit                                     & Mix of 4 and 16 bit                           & 26.62 - 94.21                             & 3.81 - 4.00        & 13.59 - 14.17  & 7.24 - 7.40 \\ \hline
\end{tabular}
\caption{Comparison of perplexity and model size between homogeneously quantized networks and our expert-only partial quantization Approach}
\label{comparison}
\end{table*}

\section{Conclusion}
In this paper, we propose an adaptive serving approach for efficient deployment of large MoE models in dynamic single-GPU constrained settings. Our approach enhances performance by reducing the overall memory footprint of MoE models through partial expert quantization. By employing partial expert quantization, our approach offers a fine-grained configuration space, allowing precise control over both throughput and output quality. Therefore, as user-defined constraints and available resources change over time, our adaptive serving approach is able to reconfigure the deployment plan and adjust to the new settings. We evaluate our adaptive serving approach on an NVIDIA A100 GPU using a Mixtral 8x7B MoE model for three language modeling benchmarks. Our results show that under GPU memory constraints of 26.28 to 53.03 GB, a throughput of 0.63 to 13.00 tokens per second is achievable. We also demonstrate that, in terms of output quality, keeping the non-expert layers in 16-bit and partially quantizing the experts to 4-bit results in perplexity ranges of 3.81 to 4.00, 13.59 to 14.17, and 7.24 to 7.40 for the WikiText2, PTB, and C4 datasets, respectively.


\bibliographystyle{unsrt}  
\bibliography{references}

\end{document}